# Phases of methodological research in biostatistics – building the evidence base for new methods

Georg Heinze[1,*], Anne-Laure Boulesteix[2], Michael Kammer[1,3], Tim P. Morris[4], Ian R White[4]

| 1 | Section for Clinical Biometrics |
| --- | --- |
|   | Center for Medical Statistics, Informatics and Intelligent Systems |
|   | Medical University of Vienna |
|   | Vienna, Austria |
| 2 | Institute for Medical Information Processing, Biometry and Epidemiology, LMU Munich, Germany |
| 3 | Division of Nephrology |
|   | Department of Medicine III |
|   | Medical University of Vienna |
|   | Vienna, Austria |
| 4 | MRC Clinical Trials Unit at UCL |
|   | Institute of Clinical Trials & Methodology |
|   | UCL, London, UK |

\* Correspondence to:

Georg Heinze

Medical University of Vienna

Center for Medical Statistics, Informatics and Intelligent Systems

Section for Clinical Biometrics

Spitalgasse 23

1090 Vienna

Austria

e-mail: georg.heinze@meduniwien.ac.at

# Abstract

Although the biostatistical scientific literature publishes new methods at a very high rate, many of these developments are not trustworthy enough to be adopted by the scientific community. We propose a framework to think about how a piece of methodological work contributes to the evidence base for a method. Similarly to the well-known phases of clinical research in drug development, we define four phases of methodological research. These four phases cover (I) providing logical reasoning and proofs, (II) providing empirical evidence, first in a narrow target setting, then (III) in an extended range of settings and for various outcomes, accompanied by appropriate application examples, and (IV) investigations that establish a method as sufficiently well-understood to know when it is preferred over others and when it is not. We provide basic definitions of the four phases but acknowledge that more work is needed to facilitate unambiguous classification of studies into phases. Methodological developments that have undergone all four proposed phases are still rare, but we give two examples with references. Our concept rebalances the emphasis to studies in phase III and IV, *i.e.*, carefully planned methods comparison studies and studies that explore the empirical properties of existing methods in a wider range of problems.

# Introduction

Career plans and funding calls in biostatistical methodology often revolve around 'novelty' and 'innovation'. This stimulates the development of new methods, and leads to results that show a new method working well. For example, asymptotic properties of a method are established and the finite-sample case investigated using simulation studies. Recently it was demonstrated that using simulation studies, 'new' methods can easily be proven to be optimal [1]. A paper introducing a new method and demonstrating its superiority over existing methods with simulation studies should therefore be treated with caution; we must be aware that these simulations may be prone to inventor bias [2]. Narrow asymptotic results and simulation studies may not create a sufficiently broad base of evidence to ensure the trustworthiness of that method. While new methods are essential to solve existing problems, users of methods need to understand which methods work well when. A trustworthy method keeps its essential operational characteristics in a wide variety of settings where it might be applied, or is sufficiently well understood such that a user of the method would know when to use the method and when to avoid. More and more new methods are proposed without ever being fully investigated and adequately compared in a wider variety of situations. This creates the problem that, even though there is a plethora of methods available to the analyst, many of them are not trustworthy enough to be used in practical analyses. In order to improve this unfortunate situation we propose a framework to think about how a piece of work contributes to the evidence base. A much needed side-effect is that such a concept gives more gravitas to carefully planned methods comparison studies and to studies that explore the empirical properties of existing methods in a wider range of problems [3].

# Learning from drug development

In drug development the concept of *phases of research* was defined decades ago [4]. As research progresses from one phase to the next, many candidate treatments are dismissed because of intolerability (phase I), lack of safety or of efficacy (phase II), or ineffectiveness when compared to a placebo or standard of care (phase III). After licensing of a drug, a phase IV trial investigates long term effects and effectiveness in the real world; this may allow identification of, e.g., an expanded safety profile, potential off-label uses or treatment effect heterogeneity.

We argue that a concept of 'phases with well-defined aims' helps to build the evidence base for methodological research.

The aim of methodological research is to give applied researchers methods to obtain accurate answers to relevant questions (and to identify methods that fail to do this), and the necessary understanding to use the methods properly. Similarly, the aim of drug development is to precisely estimate a drug's beneficial and adverse causal effects in various potential application areas. In drug development, regulatory involvement ensures that the development phases achieve these aims, being efficient with early pulling out for unpromising drugs. In methodological research, just as with drug development, new methods can be worse than existing ones or have unexpected properties in some situations. For promising methods, it is not just a matter of introducing the method and getting it used; similarly to a drug, it needs to be carefully evaluated broadly in a way that onlookers can trust.

## Introducing a framework of phases of methodological research

Methodological phase I may try to prove that a method is valid from a theoretical point of view and has the potential to improve on existing methods, or may constitute the first solution to a particular problem. It may provide the necessary logical reasoning, proofs and investigation of asymptotic properties such as consistency or normality etc.. This does not mean that proofs and investigations of asymptotic properties may not be needed in later phases. In practice, such studies may often reveal no or only a small benefit of a new method and researchers often do not try to disseminate them or, if they do, journals may be reluctant to publish them. We claim that such "negative" studies, if they can explain why a method does not work in a specific situation, are needed to increase the community's understanding, to stimulate further research and to stop other researchers from investing time and resources in the same dead-end idea [5].

Methodological phase II may have the aim to prove that a method can be used with caution in an applied setting which is not completely identical to the developer's target setting. A phase II study may provide empirical evidence to demonstrate validity using simulation studies with a limited set of scenarios, or by illustrative data analyses. When browsing the table-of-contents of typical biostatistical journals, one gets the impression that phase II study reports are abundant in the biostatistical literature (see also below). Often, a given paper includes both phases I and II contributions.

Methodological phase III may investigate how the method performs in a wide range of settings and, if applicable, for different types of outcomes. This may include empirical comparisons with any existing methods. From such studies, researchers may learn in which situations and under which assumptions the method can be safely used and performs better than or at least as well as alternative methods. This includes understanding which of the method's assumptions are critical and which are not: for example, in linear regression with large samples, the assumption of normally distributed residuals is not critical in terms of consistency of point and variance estimators. Hence, a phase III study must provide substantial evidence to demonstrate a method's validity and absolute and relative performance. It should be replicable [6] and, if possible, avoid "inventor bias" by making efforts to ensure neutral comparisons [2] or at least disclose possible biases. Furthermore, it typically includes "broad" simulations in different, practically relevant settings. Several examples involving real data would have to demonstrate how to properly apply the method in question and how to interpret its results.

Methodological phase IV should establish that a method is now suitably well-understood, i.e., we know when it is the preferred method and when it is not. A phase IV study is based on extensive experience with the method. For example, a phase IV study may systematically review applications of a method, or may deal with applying the method in new settings that were not considered initially. In this phase, pitfalls of a method could be highlighted, *i.e.*, things that are likely to go wrong if the method is not applied carefully enough by a user. Likewise, a phase IV study may propose essential, practically useful diagnostics that help a data analyst to assess if any critical assumptions of a method were violated for the data in hand. Using simulations, new mathematical results [7] and example analyses of interesting case studies beyond previous applications of the method, it may identify "sweet spots" and "breakdown scenarios" for a method (in analogy to optimal use and long-term side effects of a drug). Breakdown scenarios in which the method gives suboptimal results may not have been obvious when the method was introduced and may only be discovered through extensive experience, and they may give rise to modifications and further developments. A modification may make the method applicable in further settings, and may undergo another phase III. It may also turn out that the modification is suitable only for very special target settings, and hence the modification would still be stuck in phase II. In some areas, such as machine learning, "adversarial examples", i.e., analysis situations or data sets where a method fails, are frequently published, and often stimulate research towards robustifying existing methods. In biostatistics this is still not the case, or such examples are hidden in phase II studies intended to motivate the need for another method. Nevertheless, empirical studies on the breakdown of a method, particularly if they contain explanations of why a breakdown happens, will increase our understanding, prevent others from wasting their time on it, and are therefore worth publishing.

## Examples

For a given method, it is still unusual to have all four phases of methodological research represented in publications. As positive exceptions, we describe two developments in Table 1 representing some of the authors' interests.

Table 1: Phases of methodological research with two examples. Note that the references do not represent *all* of the work done for that method at that phase (there are many articles for some of the phases).

| Phase | Goal | Example 1: Firth correction | Example 2: multiple imputation |
| --- | --- | --- | --- |
| I | Demonstrate validity, show potential to improve on existing methods, provide first solution to a problem. | 'Bias reduction of maximum likelihood estimates' [8] | 'Multiple imputations in sample surveys – a phenomological Bayesian approach to nonresponse' [9] 'Multiple imputation for nonresponse in surveys' [10] |
| II | Method can be used with caution in an applied setting. | 'A solution to the problem of separation in logistic regression' [11] | 'Multiple imputation of missing blood pressure covariates in survival analysis' [12] |
| III | Method can be safely used in a wide variety of settings. | 'No rationale for 1 variable per 10 events criterion for binary logistic regression analysis' [13], 'Firth's logistic regression with rare events: accurate effect estimates and predictions?' [14] | 'Multiple imputation of discrete and continuous data by fully conditional specification' [15] |
| IV | Method is understood well enough to know when it is and when it is not the preferred method. | 'Separation in logistic regression: causes, consequences and control' [16] | 'Multiple imputation for missing data in epidemiological and clinical research: potential and pitfalls' [17] |

In a pilot evaluation, we analysed a volume of each of four biostatistical journals. The evaluation revealed that most articles of an issue of *Biometrika* dealt with phase I studies, while phase II dominated in *Biometrical Journal*, *Statistics in Medicine* and *Statistical Methods in Medical Research*. Overall, only a few papers were found that could be classified as phase IV. The protocol and detailed results of the pilot study can be found in the supplemental material.

A reader of a scientific communication on a statistical method should be made aware of the setting for which the method was investigated. The setting could be described with a table of assumptions and descriptions of the range of problems that it comprises. The reason for differences in decisions made in phase III or phase IV studies and in their interpretation may be rooted in different sets of assumptions. We argue that clarifying these conditions will increase the reproducibility and trustworthiness of publications concerning methods development.

## Further steps

Our proposal aims to provide a framework to communicate about the development stage of a method, *i.e.*, to understand the limitations of current research, and what further work would be necessary for wide understanding and application of a method. So far we have sketched how the phases of methodological research could be defined, but one may also think of phases when developing software packages of complex methods. In order to define phases such that they are useful and practical for the scientific community, systematic assessments of methodological papers in different biostatistical journals, and a Delphi process aiming at reaching an agreement on the definitions, are probably needed. After such work, a tool could be developed that enables a methodologist to assess and specify the phase of their research. The technology readiness level calculator of NASA may be template for such a tool [18]. The wide adoption of such a framework may facilitate efficient communication, and increase trustworthiness in the methodological development process, which is to the benefit of scientific communication, *i.e.*, it helps authors, journal editors, readers and reviewers of manuscripts and grants. The ultimate goal is to ensure that users of methods are equipped with a solid evidence base that allows them to choose the appropriate method for a given challenge.

Transparently labelled methodological studies of whatever phase may also stimulate other biostatistical researchers to get interested in a method or a methodological problem and may encourage them to conduct a study in the next phase. For example, good phase IV studies which show shortcomings of existing methods can help focus thought on solutions, and so may lead to "inventing" better methods.

Research in all four phases is important for scientific advancement, but currently there are many obstacles to achieving an appropriate balance. First, many funding agencies are inclined to fund only early phase methods research, and many biostatistical journals favour papers on new methods over articles comparing existing methods. Similarly, early career methodologists are often pushed to publish 'original' research in order to get tenure. However, the classical definition of "originality" is a very narrow one. If this same standard were applied to funding for randomised trials, we would not have studies like Recovery [19] or Stampede [20]. We claim that phase III and IV methodological studies are undervalued in the statistical community and often downplayed as "yet another simulation study" or "application", yet planning and conducting properly designed studies covering a broad variety of clearly defined settings is not trivial. A successful phase III or phase IV study involves careful consideration of the practical impact of the methods, selection of reasonable simulation setups, identification of relevant example datasets, and a good working understanding of the methods themselves. Such studies provide novelty by increasing the scientific evidence base for methods and extending the scope of their safe applicability and hence are as important to scientific advancement as "inventing" a new statistical method. The acceptance of phase III and phase IV studies could be increased by introducing new ideas related to the design and conduct of simulation and comparison studies towards the same rigour as clinical trials. This includes, e.g., publishing a protocol before conducting the study, or distributing the roles of data generator, data analyst and performance evaluator between different parties. Furthermore, inventor bias should be identified and avoided or at least disclosed [21, 22].

Concerning the reporting of such studies, the main task is to transparently clarify the level of trustworthiness of methods, both in absolute and comparative terms. The biostatistical community would benefit from phase IV studies, especially when these studies clarify when a method can – rather than cannot – be recommended for the task at hand. Experienced applied statisticians may to some extent develop a good gut feeling for this difficult task, but phase IV studies would provide the objective evidence for this intuition and aid decision making for less experienced researchers.

# Conclusion

We believe a framework such as the one outlined in this commentary may make method development more trustworthy, provide an efficient tool to communicate about methods' applicability, and increase visibility of research concerned with making the applications of methods safe and successful.

# Acknowledgements/funding

IRW and TPM were supported by the Medical Research Council [grant number MC_UU_00004/07].

Phases of methodological research in biostatistics – building the evidence base for new methods

Georg Heinze, Anne-Laure Boulesteix, Michael Kammer, Tim P. Morris, Ian R. White

# Supplementary material: Representation of phases of methodological research in the biostatistical literature – a pilot study

## Aim

The aim of this pilot study was to get a first idea of the prevalences of different phases of methodological research in the biostatistical literature.

## Method

The evaluation was based on a preliminary definition of the phases of methodological research as proposed by I.W. and agreed on in June 2021 between all authors:

- Phase I: Method is valid and has the potential to improve on existing methods.
- Phase II: Method can be used with caution in an applied setting.
- Phase III: Method is understood well enough that it can be safely used in a range of settings. This includes knowing how to check its assumptions and understanding which assumptions are critical and which are not.
- Phase IV: Method is understood well enough that it is known when it is and when it is not the preferred method.

The following four biostatistical journals were chosen: Biometrical Journal, Statistics in Medicine, Statistical Methods in Medical Research, Biometrika.

Four authors each chose a journal, selected a recent issue and assigned each article of that issue to one of the four phases. The chosen issues were:

| Journal | Issue | Number of articles | Assessed by |
| --- | --- | --- | --- |
| Biometrical Journal | Volume 63, Issue 4 (April 2021) | 12 | G.H. |
| Statistics in Medicine | Volume 40, Issue 15 (July 2021) | 12 | A.L.B. |

| Statistical Methods in Medical Research | Volume 30, Issue 5 (May 2021) | 13 | M.K. |
| Biometrika | Volume 108, Issue 2 (June 2021) | 16 | T.M. |

## Results

The following table summarizes the results of the pilot evaluation:

| Phase | Biometrical Journal | Statistics in Medicine | Statistical Methods in Medical Research | Biometrika |
|---|---|---|---|---|
| 0 (pre I) | | | | 3* |
| I | 0 | 2 | | 9** |
| I or II | 2 | | 3 | |
| II | 7 | 8 | 7 | 4 |
| II or III | | | 1 | |
| III | 0 | 2 | 2 | |
| IV | 1 | | | 2** |
| Not classifiable | 1 | | | |

\* 3 papers were rated as rather theoretical with no application in mind.

\*\* 2 papers were rated as mix of phases I and IV.

## Further remarks

Feedback of the raters on the evaluation process was collected via e-mail exchange. This feedback led to the following additional remarks on the classification task:

- Generally, all raters reported that the classification task was difficult, in particular if the methods presented in a paper were outside of the specific expertise of a rater
- The definition of the phases was not yet precise enough to allow for a clear classification, e.g., between phases I and II or between II and III
- Some papers dealt with several methods. For some of the methods, those papers could have been classified as phase III, but then also some new methods were introduced.
- Some non-standard cases were found, e.g., a paper on an efficient implementation of a known method, a paper focusing on a new, more efficient implementation of a known method, and a paper extensively comparing several new methods for a problem without existing method.
- There were even papers which could be classified as a mix of phases I and IV.
- Some papers were not classifiable (because they did not deal with methods) or they presented theoretical work for which a corresponding application was not obvious.

## Conclusion

The authors concluded that a more unambiguous classification would need:

- more precise definitions of the four phases
- a specially designed tool to facilitate the rating
- clear rules how to deal with papers that deal with multiple methods in different phases
- to minimize variation between raters, a paper would have to be assessed by more than one person